# Performance assessment of medical and non-medical CPAP interfaces used during the COVID-19 pandemic

Marco Marini[1], Lorenzo Capponi[2], Ilaria Battistoni[1], Tommaso Tocci[3], Giulio Tribbiani[3], Francesco Merante[4], Matteo Francioni[1], Maria Vittoria Matassini[1], Giulia Pongetti[1], Matilda Skhoza[1], Luca Angelini[1], Roberto Marsili[3], Marco Mazzanti[5], Gian Piero Perna[1], Gianluca Rossi[3], Alberto Gili[1]

[1] *Intensive Cardiac Care Unit and Cardiology, Department of Cardiovascular Sciences, Ospedali Riuniti, Ancona, Italy*
[2] *Department of Aerospace Engineering, University of Illinois Urbana-Champaign, Urbana, US*
[3] *Department of Engineering, University of Perugia, Perugia, Italy*
[4] *San Giovanni Battista Pulmonology Unit, Foligno, Italy*
[5] *Department of Cardiology, Barts Heart Centre, NHS Foundation Trust, London, UK*

ABSTRACT

Background: At the beginning of 2020, a high number of COVID-19 cases affected Italy in a short period, causing a difficult supply of medical equipment. To deal with the problem, many healthcare operators readapted different masks to medical devices, but no experiment was conducted to evaluate their performance. The aims of our study were: to assess the performances of three masks and a CPAP helmet in their original configuration and after modifications, in the maintenance of mean pressure and half-amplitude variations using different PEEP valves and to analyse the impact of antibacterial (AB) or antibacterial-viral (ABV) pre-valve PEEP filters on the effective PEEP delivered to the patients. Four pressure ports were installed on each mask (three on helmet), mean values and half amplitudes of pressure were recorded. Tests were performed with any, AB, ABV filter before the PEEP valve.
CPAP helmet was the most efficient interface producing more stable mean pressure and less prominent half-amplitude variations but the non-medical masks, especially after the modifications, maintained a stable mean pressure value with only a moderate increase of half-amplitude. The use of AB and ABV filters, produced respectively an increase of ± 2,23% and 16.5% in mean pressure, compared to no filter condition. CPAP helmet is the most reliable interface in terms of detected performance, but readapted masks can assure almost a similar performance. The use of ABV filters before the PEEP valve significantly increases the detected mean pressure while the AB filters have almost a neutral effect.

Funding:
Corresponding author: Roberto Marsili, email: roberto.marsili@unipg.it

## 1. INTRODUCTION

In December 2019, severe acute respiratory syndrome coronavirus 2 (SARS-CoV2) emerged in Wuhan, China and rapidly spread all over the world. [1] In Italy the first infection was detected on February 21th 2020, and the country was significantly affected with more than 230.000 new cases and 30.000 deaths in the following three months. [2] The abrupt diffusion of COVID-19 in Italy in such a short timeframe, caused a sort of "infective Tsunami" that asked for extraordinary measures by the National Health System, especially in terms of supply of ventilation devices. In fact, from the analysis of different registries, about 50% of all patients admitted for SARS-CoV2 required the use of different non-invasive ventilation techniques (NIV) or invasive mechanical ventilation (IMV), while only 3% needed a more advanced support such as Veno-Venous Extra Corporeal Membrane Oxygenation (VV-ECMO) [3-5]. The use of NIV in the management of acute respiratory distress syndrome (ARDS) has always been a matter of debate especially because the subpopulation of patients who might benefit remains unclear [6]. Moreover, doubts have significantly increased after the first data obtained from COVID-19 patient analysis, that show some peculiars features of ARDS [7-9].
At present time there are no clear recommendations on which ventilatory modality should be the first one in the treatment of COVID-19 patients with respiratory insufficiency and although the use of Continuous Positive Airway Pressure (CPAP) or NIV with pressure support (PS) are recommended by many organizations [10-12], others advocate for high flow nasal cannulas (HFNC) as first choice [13] or do not recommend the use of NIV [14]. An exhaustive knowledge of every component of the ventilatory system, a frequent measurement of positive end-expiratory pressure (PEEP) and fraction of inspired oxygen (FIO2) delivered as well as a well-trained healthcare teams, are crucial to guarantee the effectiveness of NIV techniques [15]. In particular, CPAP masks have to be well tolerated by patients, maintain high adherence and sealing characteristics and be able to preserve a constant PEEP value during their utilization. Due to the emergency situation and the poor supply of medical equipment, many healthcare operators collected different masks (not only for medical use) and transformed them into medical devices for NIV [16]. Unfortunately, no specific data exist about these masks' performance in terms of tolerance and efficacy on maintaining a stable pressure. mean pressure inside the device and half-amplitude. The aim of our study was to evaluate the mean pressure and the half-amplitude variations inside three masks readapted as CPAP masks (two full face snorkeling masks and one full-face medical masks for NIV) and one CPAP helmet, using different PEEP values and reassess the results after some structural modifications on the masks. In addition, we also tested how the use of antibacterial (AB) or antibacterial and antiviral (ABV) pre-valve PEEP filters could affect the final mean pressure inside the ventilation devices.

1. **MATERIALS AND METHODS**

1.1. **Sensor calibration and dynamic characterization of connecting pipe**

In this research, a high resolution piezoresistive pressure sensor was used. The static sensor calibration was carried out by means of a classical approach [17]. The uncertainty $S_{qi}$ of the sensor, that was used in this research, is then evaluated on the full scale (*f.s.*) as: $S_{qi} = 0.037 \% \ f.s.$. To properly protect the sensor from humidity, it was connected to the mask using a common polyurethane pipe. Since, the length of the pipe which connects the pressure sensor to the measurement port can affect the dynamic measurement chain performances as already described [18], simpler and more practical models were recently proposed [17-19-20]. As usually performed in dynamic pressure measurement for experimental fluid dynamic studies the frequency content of the measurand is first of all estimated and compared to the frequency response function of the measurement chain (connecting tube and sensor) [17]. The producer declared frequency response of the sensing element is up to 1000 Hz, so no dynamic distortion is expected in the frequency content typical of the human inspiration end expiration cycle. To measure the frequency response function of the connecting tube the setup shown in Figure 1 was prepared.

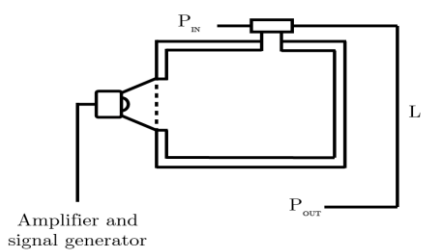

Figure 1 - Dynamic characterization setup (Pin: Input pressure; Pout: Output pressure; L: Tube length)

The length of the connecting pipe was determined after measuring its frequency response function with a setup consisting of a harmonic signal generator, a signal amplifier and two pressure sensors to measure the input (Pin) and output (Pout) pressures of the second pressure sensor connected through the tube. Different length of polyurethane pipes (1 m, 5 m and 10 m) was tested obtaining various frequency response function amplitude. A dynamic signal amplitude change can be noticed for 5 m and 10 m tube length for frequencies higher than 4 Hz. In addition to that, under the oversight of a medical team, four various patterns of breath (i.e., a normal, dyspnea, tachypnea and breathing during Muller manoeuver) simulated by a single healthy human tester were measured. As shown in Figure 2a, the frequency content of the measurand is in the 0 – 4 Hz range in all four respiration patterns. In addition, a dominant frequency corresponding to the breath-in and the breath-out ventilation period is observable mostly within 2 Hz and no significant harmonics have been found in spectra. This is why we decided to use the 1 m pipe length in the further analyses.

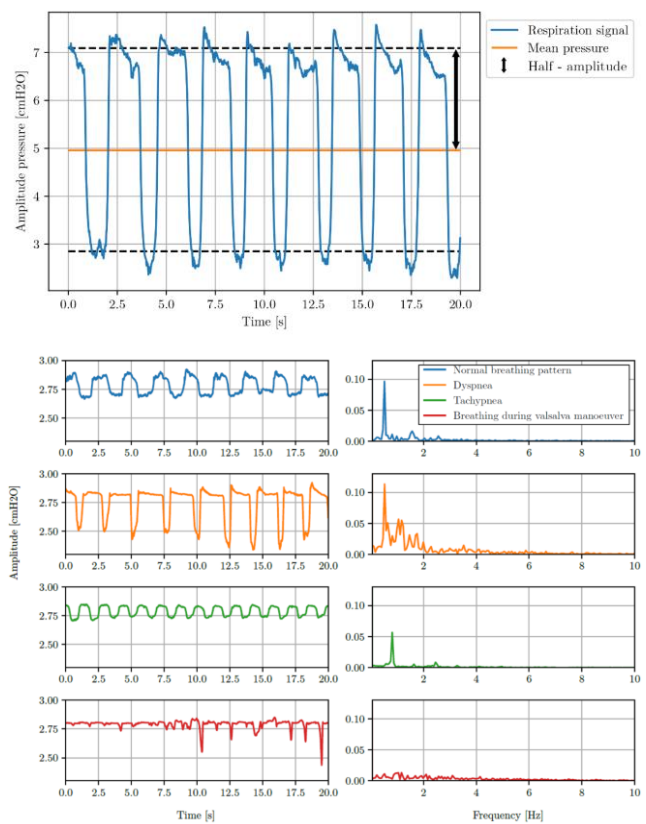

Figure 2 – (a) Mean and half-amplitude pressure definition considering a standard respiration signal (b) Simulated respirations signals time histories and spectra; Each coloured time-frequency combination stands for each tested simulated respiration

1.2. **Performances of full-face masks**

The mean pressure and stability of the air pressure inside a ventilation device are import parameter that properly describe the performances of a CPAP mask and should be longitudinally verified during the ventilation process. In fact, the particular mask configuration such as filters or input-output breathing conduits position, influences the mean and amplitude pressure during the ventilation process (Figure 2b). Due to this, firstly, the influence of output filters and, later, of the morphology of the input output breathing configuration, were researched. The experimental test campaign was carried out in order to compare the performances of four different ventilation interfaces: two snorkeling adapted masks, Mares Sea Vu Dry (mask M1) and, Decathlon Easybreath (Mask M2), a NIV medical mask, Pulmodyne - Bitrac select SE medical (mask M3) and Dimar CPAP helmet (mask M4) (Figure 3). The masks were firstly tested in their original configuration (ORC), with PEEP 5 cmH2O, supplied by a common laboratory centrifugal compressor, connected to the masks, which provided a dry gas mixture. Four pressure ports were installed on each mask, while only three ports on M4 helmet, and the tests were repeated using three different filter configurations: without any filter at the output (NF), with an Intersurgical Flow-Gard-Breathing antibacterial filter (AB), and with a DAR HME small electrostatic antibacterial and antiviral filter (ABV). The helmet M4 was tested using an input-reservoir (DimAir 25 L) as in a standard configuration. Adapting junctions and pressure ports were realized by additive manufacturing technologies.

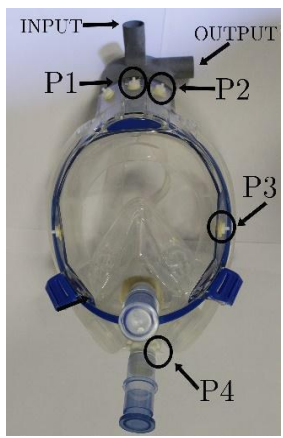

Figure 3 - Tested masks with installed pressure ports (P1-P4) in the original configuration (ORC) and in the modified configuration (MOC). This is the figure 3 legend: a) M1 mask; b) M2 mask; c) M3 mask d) M4 helmet; f) M1 mask; e) M3 mask; INPUT: 99 entrance O2/air OUTPUT: air outlet where to attach PEEP

Once defined the influence of output filters at the same pressure value, the structural configuration of masks M1 and M3 was modified (MOC) (fig 3). A second hole for PEEP valve application in M1 mask was created, moving the breathing-in conduit close to the respiration area, in front of the mouth. For M3 mask, a 3D printed watertight adapter was designed. The experiments were repeated in MOC, in no filter condition, using progressive increased PEEP value (5 cmH2O –108 7.5 cmH2O - 10 cmH2O – 12.5 cmH2O). Intersurgical leaf spring peep valves were used for 5 and 7.5 cmH2O, while Harol linear spring valves were used for 10 and 12.5 cmH2O. For the sake of repetition, the experiments were conducted on 3 different healthy humans, who repeated 5 times each experiment, for a total of 15 measurements under the same test configuration. The time duration of each experiment was 20 seconds.

### 1.3. Statistical Analysis

The proposed statistical approach is shown in Figure 8. The uncertainty analysis (UNI4546) was performed, at the 68% of confidential level, by considering all the 15 experiments under the same PEEP-filter combination, in order to obtain mean and half-amplitudes values of each 15-tests distribution. Thus, the total uncertainty is represented as an ellipse, which has for the x- and y-axis the uncertainty of the mean values and half-amplitudes values, respectively. The approach is planned in this way: at the end of all the experiments, the uncertainty of each experiment is evaluated as shown in Figure 4 and compared to the other in order to identify the worst one, that we consider conservatively as the global uncertainty. Finally, the linearity of the results was also researched by computing the Pearson coefficient during the results interpolation.

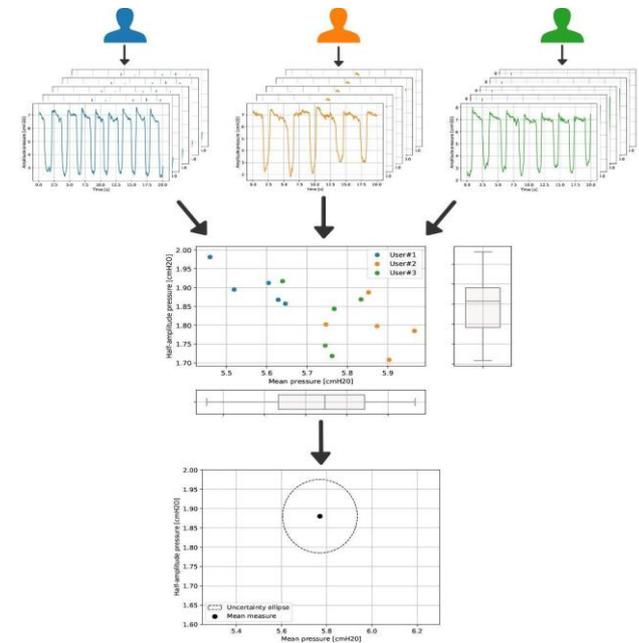

Figure 4 - Proposed statistical and experimental approach: sample data under the same PEEP-filter combination

## 2. RESULTS

Analysing the behaviour of the pressure inside each mask by measuring the mean value and the half-amplitude at the four ports, presented in Figure 3, it was found that all the masks showed similar trend from the input (P1 port) and output (P2 port) conduits and, moreover, inside the mask (P3 and P4 ports): as expected, the pressure was higher in the intake conduit, it decreased inside the mask and it was finally lower in the exhaust conduit. However, the M2 mask shows half-amplitude pressure higher than the M1 mask and a higher reduction of the mean pressure in the output conduit. On the other hand, the M3 mask shows more stable results both in terms of mean and half-amplitude values of pressure. In addition, it was found that the pressure measurements in P3 port gave more reliable information of the pressure inside each mask, for this reason, only P3 measurement was taken into account and, moreover, by considering the similarity between the masks M1 and M2 in P3 port in no-filter configuration, only M1 results will be shown. In Table 1 the mean values and the half-amplitude measurements considering M1, M3 and M4 masks on the pressure port P3, in ORC and MOC at PEEP values of 5, 7.5, 10 and 12.5 cmH2O in NF condition are shown. M4 helmet provides the lowest half-amplitude pressure compared to the other masks, irrespectively of PEEP values. The differences in mean pressure measured using different PEEP values with ORC and MOC configurations are negligible for M1 and M3, while the half-amplitude values are consistently higher for M3 mask, and increase using higher PEEP levels especially in the ORC.

Table 1 - Mean value and standard deviation of pressure measurements of M1, M3 and M4 masks. This is the Table 1 legend: ORC: original configuration; MOC: modified configuration; Half-ampl: half amplitude pressure

| Test case | | M1 | | M3 | | M4 | |
|---|---|---|---|---|---|---|---|
| | | Mean | Half-ampl | Mean | Half-ampl | Mean | Half-ampl |
| ORC | 5 | 5.651 | 2.044 | 6.354 | 5.058 | 6.083 | 1.881 |
| | 7.5 | 7.19 | 2.344 | 8.899 | 5.133 | 8.94 | 2.712 |
| | 10 | 10.872 | 3.415 | 10.619 | 7.227 | 10.823 | 3.421 |
| | 12.5 | 13.028 | 3.121 | 12.962 | 8.367 | 13.522 | 3.691 |
| MOC | 5 | 5.399 | 2.213 | 6.03 | 2.617 | - | - |
| | 7.5 | 8.565 | 2.417 | 8.464 | 3.095 | - | - |
| | 10 | 11.145 | 2.70 | 9.677 | 3.795 | - | - |
| | 12.5 | 12.451 | 2.196 | 13.074 | 4.533 | - | - |

In Figure 5 the linearity of the results is remarked, where the coloured ranges around the lines show the global uncertainty on the half-amplitude. In each line, except the intersection points, no overlaps occur. Due to this, each measurement can be considered statistically significant. For both M1 and M3 masks, the pressure fluctuation in the modified configuration is found to be lower than in the original configuration and closer to the M4 helmet values. In order to enhance the linearity of the results, the Pearson coefficient R for the measured values distribution was computed. As shown in Table 2, the coefficient is close to the unity for all the test cases, and it proves the linearity of the measurements.

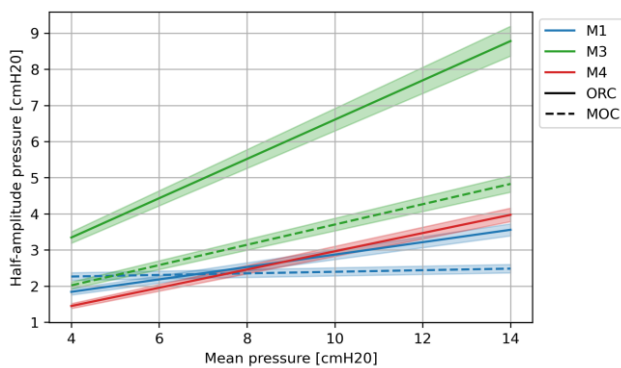

Figure 5 - Linear interpolation of measurement results (ORC: original configuration; MOC: modified configuration)

Table 2 - Pearson correlation coefficients R (ORC: original configuration; MOC: modified configuration)

| Test case | | R |
|---|---|---|
| ORC | M1 | 0.899 |
| | M3 | 0.931 |
| | M4 | 0.975 |
| MOC | M1 | 0.889 |
| | M3 | 0.984 |

As shown in Figure 6, the introduction of the AB filter causes a variation of mean pressure of ± 2,23% compared to NF condition, while an increasing of mean pressure of about +16,5% was measured when the ABV filter was used during the measurements. No particular influence on the half-amplitude was observed respect to the used filter, but mostly on the type of the used mask. From the statistical analysis, by considering all the masks configurations, the worst results in terms of uncertainty were, for the mean pressure, about 0.17 cmH2O (2.8% of the mean value), while for the half-amplitude was 0.9 cmH2O (4.7% of the mean value of the half amplitude). These results are thus to be considered valid as the global uncertainty for all the performed experiments (not to be confused with the statistical significance of the results). Moreover, considering the results obtained and shown in Figure 10, where the influence of the filters was research in terms of pressure mean value increasing, and considering the global uncertainty, it was found that the results obtained from NF-AB comparison of ± 2,23% of the mean value cannot be considered significant due to the mean pressure global uncertainty of the 2.8% (obtained from the statistical analysis), that does not allow to have two separated uncertainty-ranges. On the contrary, the uncertainty evaluated considering the ABV filter can be considered statistically significant, being +16.5% of the mean value.

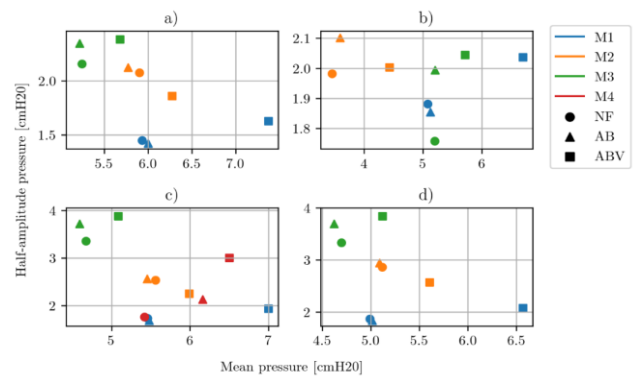

Figure 6 - Mean values and half-amplitudes of pressure measurements considering M1, M2, M3 masks and M4 helmet, in three different output-conditions at PEEP 5 cmH2O (NF = No-filter, AB = Antibacterial filter, ABV = Antibacterial and antiviral filter): a) Pressure port P1; b) Pressure port P2; c) Pressure port P3; d) Pressure port P4

3. DISCUSSION

The study shows that although CPAP helmet represents the most reliable interface in terms of mean pressure and half-amplitude. Likewise, medical and non-medical masks readapted for non-invasive ventilation demonstrate a fairly good performance. In particular, the "non-medical" masks (M1 and M2) compared to M3 are more effective, providing a reliable mean pressure value and low variable half-amplitude pressure. This efficiency remains constant for all PEEP values from 5 cmH20 to 12.5 cmH20 and it is not particularly affected by the type of configuration, producing negligible differences between the standard and modified configuration. On the contrary, M3 guarantees less stable mean pressure, especially if used in standard configuration with a T-tube causing an excessive excursion of pressure with values of almost 208 9 cmH20. Using the modified configuration, the interference with air / O2 input has been reduced and the results obtained have been comparable to those obtained with M1 mask. The different results obtained with M1 and M3 masks can be explained by the divergence in terms of material (softer silicon for the full-face snorkeling masks compared to a stiffer one in the full face NIV mask) and support surface (bigger and more comfortable in snorkeling mask, smaller and less fitting in NIV mask). In fact, the NIV performance is mainly conditioned by interface efficiency and patient's tolerance to it [20-21], Although full-face masks for CPAP are undoubtedly much more comfortable than nasal or gold-nasal masks, especially in prolonged time use [22], the CPAP helmet remains the most popular choice for patients due to its high tolerance [23]. Moreover, in a previous experience [24], in ARDS patients, the use of NIV helmet compared to NIV face mask, was associated to a better functional independence at one year after hospital discharge. As already mentioned, the recently published guidelines did not recommend any specific method to be used as first line in COVID-19 patients [13-25] and more importantly, when it is the right time to switch to an invasive mechanical ventilation [26]. However, it has been established, and confirmed by a meta-analysis by Ferreyro [27], that the use of non-invasive oxygenation in adult patients with acute hypoxemic respiratory failure, compared to standard ventilation therapy is associated with a lower risk of death. A recent metanalysis on the role of NIV in the treatment of acute lung injury and ARDS found an overall reduction in intubation without a reduction of mortality. The authors conclude than NIV should be use cautiously in these patients using a strict monitoring [28-30]. From the analysis of our experiments, we clearly observed an increase, not statistically significant, in pressure mean value due to AB filter use while the ABV use is associated to a significant increase of +16.5% of the mean pressure value. Moreover, how filters could affect the mean values during the hours is unknown. The use of filters before the PEEP is highly recommended by all guidelines [12-14] in order to reduce the aerosolization and the virus spread. As easily foreseeable, these filters, increase the pressure inside the system so as to determine a higher 233 PEEP than set causing an obstacle to the air outlet, as reported in previous experiences in mechanically ventilated patients [31]. The increase in pressure is linked to the type of filter, higher in the antiviral filters and lower in the antibacterial ones, where the increase could be considered almost irrelevant. Furthermore, it must be absolutely considered that filters tend to get wet over time, due to their intrinsic characteristic, increasing their resistance and consequently the pressure reached. For all these considerations the NHS guidelines suggest a daily change of filters [25]. In order to check whether the application of AB or ABV filters provokes an excessive increase of pressure inside the masks a frequent use of manometer is advisable. This simple tool can help clinicians to verify that the pressure values inside the mask really correspond to the chosen PEEP and eventually reduce the PEEP value to avoid a dangerous increase of pressure [32]. Major limitation of the study is that we present a preliminary observational study conducted during the COVID-19 pandemic then more experiments could be advisable to confirm the results obtained. We analyzed only some specific type of masks/CPAP helmet therefore the results could not be extended to other interfaces. Moreover, all the experiments were performed on healthy volunteers so how the dyspnoeic pattern could modify the results is unknown. In conclusion our study shows that CPAP helmet remains the most reliable interface in terms of mean pressure value and half-amplitudes, but some non-medical masks and full-face NIV masks, especially with some structural modifications, could be effectively used as CPAP masks could assuring almost a similar performance. The use of ABV filters before the PEEP valve significantly increases the detected mean pressure value inside the mask while the AB filters show almost a neutral effect. A repetitive assessment of the PEEP effectively delivered is advisable in order to optimize the use of NIV and avoid harmful situations for the patients.